\pdfoutput=1
\documentclass[12pt,a4paper]{article}

\usepackage{graphicx}
\title{A heated vapor cell unit for DAVLL in atomic rubidium}

\author{Daniel J McCarron, Ifan G Hughes, Patrick Tierney\\ and Simon L Cornish\\
\small Department of Physics, Durham University, South Road\\
\small  Durham, DH1~3LE, UK\\
\small \texttt{s.l.cornish@durham.ac.uk}}

\date{\today\\
\small pacs: 32.60.+i, 32.70.Jz, 42.55.Px}

\begin{document}
\maketitle

\begin{abstract}
The design and performance of a compact heated vapor cell unit for realizing
a dichroic atomic vapor laser lock (DAVLL) for the $D_2$ transitions in atomic rubidium is
described. A 5~cm-long vapor cell is placed in a double-solenoid arrangement to produce
the required magnetic field; the heat from the solenoid is used to increase the
vapor pressure and correspondingly the DAVLL signal. We have characterized experimentally the dependence of
important features of the DAVLL signal on magnetic field and
cell temperature. For the weaker transitions both the amplitude and gradient of the signal are increased by an order of magnitude.
\end{abstract}

\section{Introduction}
Diode lasers are extensively used in atomic physics experiments, especially in the field of laser cooling
and trapping \cite{nobel97}. The active stabilization, or ``locking", of a laser's frequency is a
key feature of many such experiments.  A frequently-used scheme is the dichroic atomic vapor laser lock (DAVLL) \cite{Cheron94,Corwin98} which relies upon the differential Doppler-broadened absorption of orthogonal circular polarisations ($\sigma^{+}$ and $\sigma^{-}$) in an atomic vapor in the presence of an applied magnetic field. The magnitude of the applied field is set to give Zeeman shifts that are comparable to the Doppler-broadened width of the absorption line, so that the resulting DAVLL signal permits the lock-point to be tuned over a large range of frequencies and has a broad capture range (defined as the frequency excursion the system can tolerate and still return to the desired lock-point). Additionally, the amplitude of the DAVLL signal scales with the line-center absorption. As a consequence, the signal is adequate for the strong $^{85}$Rb $F=3 \rightarrow F^{'}$ and $^{87}$Rb $F=2 \rightarrow F^{'}$ transitions, but offers scope for an order of magnitude improvement for the weaker $^{85}$Rb $F=2 \rightarrow F^{'}$ and $^{87}$Rb $F=1 \rightarrow F^{'}$ transitions (``repump" transitions in laser-cooling experiments). We have previously demonstrated an increase in the signal amplitude and gradient using a cell heated with thermo-electric heat pumps combined with permanent magnets to provide the field~\cite{Fred07}. However it proved difficult to achieve a stable DAVLL signal owing to the transmission of heat from the cell to the magnets. In this paper we present a design for a compact heated vapor cell unit for realizing DAVLL which largely circumvents these problems by using a solenoid to both provide the required magnetic field and to heat the cell.

\section{Principles of DAVLL and apparatus}
\begin{figure}[!ht]
\centering
\includegraphics[trim = 105mm 150mm 20mm 55mm, clip, scale=0.9]{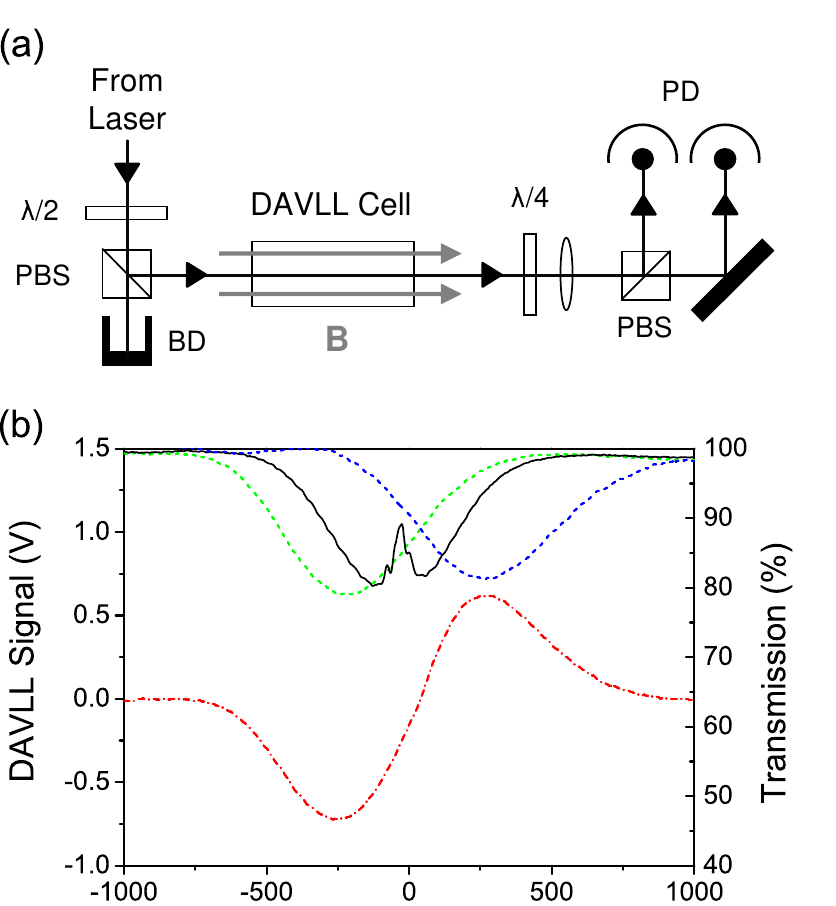}
\caption{(a) A schematic of the optical components used to generate the DAVLL signals. Polarizing beam splitters (PBS) are used to pick-off a fraction of the main beam and to analyze the signal; two photodiodes (PD) record the intensity of the orthogonally polarized beams; the majority of the laser power is caught in the beam dump (BD). The half-wave plate determines which fraction of the light is sent into the cell, the quarter wave plate converts the two opposite hand circularly polarized components into two linear orthogonally polarized beams. (b) DAVLL on the $^{85}$Rb $F=2 \rightarrow F^{'}$ transition. The left-hand scale shows the  DAVLL signal (red dot-dash),
the right-hand scale   shows the  transmission through the cell.  The intensity of the two circularly polarized beam components are shown (blue and green dashed) and the reference sub-Doppler spectrum (black solid) recorded in a separate cell for calibration. }\label{fig:fig1}
\end{figure}

The principles of DAVLL are clearly outlined in \cite{Cheron94,Corwin98} and only a summary is presented here.
A linearly polarized probe beam is incident on an atomic vapor contained in a
cell (see figure~\ref{fig:fig1}~(a)). The wavevector of the light is parallel to the axis of an applied magnetic field.
After exiting the cell the beam passes through a quarter wave plate before impinging on a polarizing beam
splitter (PBS). The linearly polarized beam incident on the cell can be decomposed into two orthogonal circularly polarized beams of equal amplitude.  The signals on the detector in the output arms of the PBS are proportional
to the intensity of the right and left circularly polarized beams. For the case of no field,
both circular polarizations are absorbed equally and the difference in signals is zero for all frequencies.
For  a finite magnetic field the degeneracy is lifted and the medium becomes dichroic.  The center of the
absorption line for one hand of circular polarization is displaced to higher laser frequency, whilst
the other absorption line is displaced to lower frequency. Consequently the difference in readings of the two photodetectors  has a dispersion-like shape, with a zero at line-center. The difference signal forms the output of the DAVLL spectrometer used to regulate the frequency of the laser (the ``error signal").

Figure~\ref{fig:fig1}~(a) shows the apparatus and (b) an experimental DAVLL signal.
The experiment used a Sacher Lasertechnik Lynx TEC 120 external cavity diode laser system. The output beam had a
$1/e^2$ radius of (0.83$\pm$0.01)~mm vertically  and (0.98$\pm$0.02)~mm
horizontally. Saturated absorption/hyperfine
pumping spectroscopy \cite{Macadam92, Smith04} was used for frequency reference; the scans were
calibrated and checked for linearity against the known rubidium hyperfine structure \cite{Arimondo77}.  The optical set-up
utilized two narrow band polarizing beam splitters (Casix PBS0101); a low-order half-wave plate
(Casix WPL1210); and a zero-order quarter-wave plate (Casix WPZ1210) for the analysis. The
half-wave plate in combination with the polarizing beam splitter was used to both set the power and improve the polarization purity of the beam in the DAVLL setup. Neutral density filters were used to set a DAVLL probe laser
power of (41.1$\pm$0.5)~$\mu$W. The DAVLL detectors
have a measured responsivity of (0.462$\pm$0.003)~A/W and (0.476$\pm$0.003)~A/W and the
amplifier a transimpedence of (0.994$\pm$0.001)M$\Omega$;
consequently a 20~$\mu$W beam generates a 9.5~Volt signal.

\section{Unit design criteria}
Our previous study of  DAVLL lineshapes \cite{Fred07}   with the $D_2$ transitions in $^{85}$Rb and $^{87}$Rb used either (i) a solenoid to generate a uniform magnetic field along the length of the  cell, or (ii) a heated cell with permanent magnets. For the former, care was  taken to collect data quickly after the solenoid was energized to ensure the experimental conditions (vapor pressure in the cell and temperature-dependent birefringence in the cell windows) were the same for all data sets.  The ``footprint" of the apparatus was large ($\sim$500~cm$^2$), dominated by the  solenoid.  For the latter, the increase in vapor pressure and concomitant  absorption yielded significantly larger signals, however the thermal stability of the apparatus was poor.  Our motivation for this work was to generate suitable locking signals for laser-cooling experiments with a compact arrangement, and to utilize the stable field and unavoidable Joule  heating from  a  solenoid to increase the vapor pressure of the cell.

Previous work shows that even an elaborate theoretical treatment of DAVLL spectra  does not wholly account for the experimental spectra obtained \cite{Reeves06}.  Hence we chose to focus our attention
on the empirical behavior of the DAVLL spectra obtained (with the starting point  that a line-center absorption of
$\sim$75\% yields the steepest gradient \cite{Fred07}).

\begin{figure}[!ht]
\centering
\includegraphics[trim = 90mm 90mm 30mm 65mm, clip, scale=1]{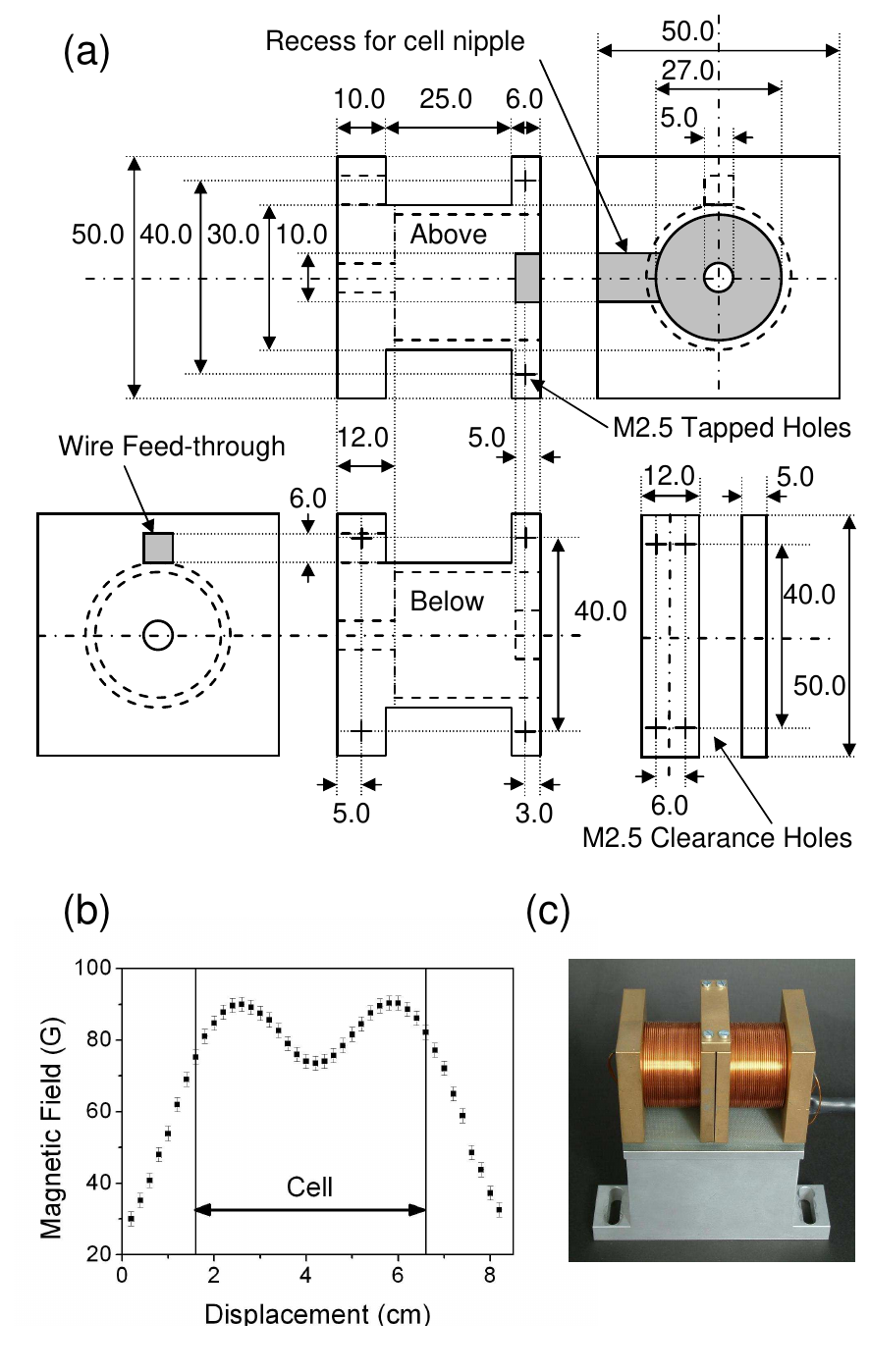}
\caption{(a) Technical drawing showing the dimensions of the unit. (b) A plot of the axial variation of the magnetic field along the unit when operated at 1.50~A and wound with six layers of wire. (c) A photograph of the assembled unit.}\label{fig:fig2}
\end{figure}

A schematic of the unit is shown in figure ~\ref{fig:fig2}~(a), and the assembled unit in (c).
The cell holder was made from brass to allow good thermal conduction to the vapor cell.  An insulator was incorporated between the cell holder and the aluminium stand.  To ensure that the rubidium did not condense on the cell windows the holes in the cell holder to allow propagation of the DAVLL beam were restricted to a diameter of 5~mm.  The separation between the cell (Newport 2010-Rb-02) and the wire was restricted to a minimum of 1.5~mm.
A small section was removed from the inner face of each cell holder in order to house the cell nipple. As the vapor pressure in the cell is limited by its coldest point it is important this hole was capped by a brass lid.

To optimize the design of the unit a model to simulate the magnetic field along the axis of the cell was developed.
This integrated the contribution to the field from each wire according to the Biot-Savart law.
The wires have a two-fold contribution: to provide the magnetic field in the cell to make the medium dichroic; and to heat the cell to increase the vapor pressure.  Thus the field and cell temperature in this design are dependent.
One way to alter the relationship between them is to alter the number of layers of wire.
Enameled copper wire of diameter 0.71~mm was used.  The choice of wire gauge is a trade off between the wire resistance
and ease of coil winding.

The axial field is shown in figure~\ref{fig:fig2}~(b); it is symmetric, with two maxima separated by approximately 4~cm.  Previous studies have shown that
there is little variation in the DAVLL signal produced by a solenoid with a mean field of 151~Gauss and a standard deviation along a 5~cm cell of 1~Gauss and that arising from permanent magnets with a mean field of 151~Gauss and a standard deviation of 59~Gauss \cite{Fred07}. However, it is important to keep the width of the part housing the cell nipple to a minimum, otherwise the field in the region between the two solenoids becomes too small and the quality of the DAVLL spectrum is significantly degraded. The unit developed here when operating at 1.50~A and wound with six layers of wire produced a mean field of 84~Gauss and a standard deviation along the cell of 6~Gauss.  We note that there exist designs \cite{Yashchuk} using  permanent magnets where the  fields are fully contained and thus  can be used in proximity of magnetically-sensitive instruments.

\section{Results and discussion}
The data reported here were collected using either six layers of copper wire, input currents of 1.00 to 2.75~A, generating an external temperature of  26~$^\circ$C to 63~$^\circ$C and a field of 56 to 154~Gauss; or two layers of wire, input currents of 1.50 to 4.50~A, generating an external temperature of  25~$^\circ$C to 58~$^\circ$C and a field of 29 to 87~Gauss.

\begin{figure}[!ht]
\centering
\includegraphics[trim = 0mm 0mm 125mm 180mm, clip, scale=1]{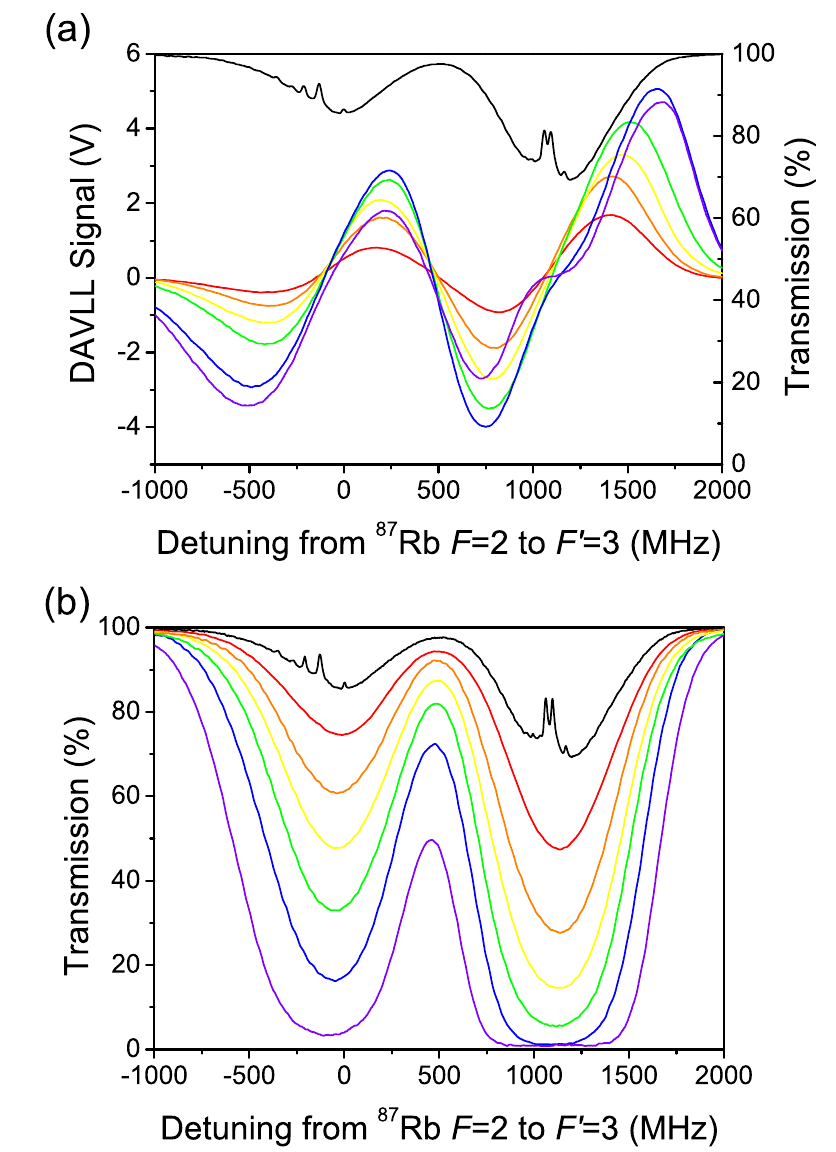}
\caption{(a) DAVLL on the $^{87}$Rb $F=2 \rightarrow F^{'}$ transition and $^{85}$Rb $F=3 \rightarrow F^{'}$ transitions, together with the sub-Doppler reference spectroscopy signal (right-hand scale). (b) The  transmission through the cell obtained by momentarily turning off the current through the solenoids, again with the reference spectroscopy signal. The data were obtained with the unit having six layers of wire; the current increases from 1.00 to 2.25~A in steps of 0.25~A.}\label{fig:fig3}
\end{figure}

Figure~\ref{fig:fig3} shows data obtained for the $^{87}$Rb $F=2 \rightarrow F^{'}$ and
$^{85}$Rb $F=3 \rightarrow F^{'}$ transitions with 6 layers of wire.
It is clear that the DAVLL signals (shown in figure~\ref{fig:fig3}~(a)) are highly dependent on the solenoid current.
The different absorption curves in figure~\ref{fig:fig3}~(b) show that with this design it is possible to increase the vapor pressure to a point where the medium becomes sufficiently optically thick that the transmission is less than 0.5\%.  There are two factors which govern the evolution of the DAVLL spectrum amplitude and line-center gradient.  As the current increases, the vapor pressure in the cell increases, as do the absorption and signal amplitude.  Higher currents give larger fields, which also increases the magnitude of the signal  initially. If the field produced is sufficiently large for the Zeeman shift to exceed the Doppler width of the line the signal amplitude decreases. The gradient also grows initially with increasing current, before reaching a maximum and then decreasing for higher currents.

\begin{figure}[!ht]
\centering
\includegraphics[trim = 0mm 0mm 125mm 180mm, clip, scale=1]{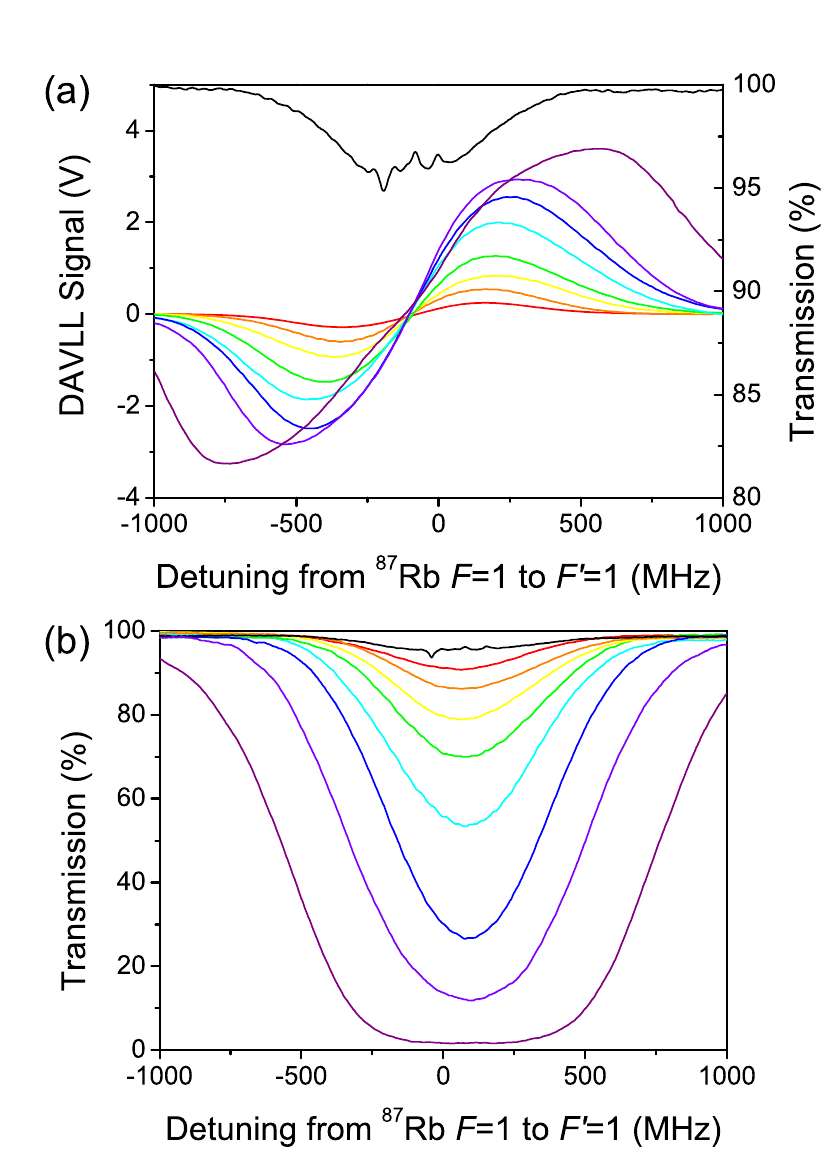}
\caption{(a) DAVLL on the $^{87}$Rb $F=1 \rightarrow F^{'}$ transitions, together with the sub-Doppler reference spectroscopy signal (right-hand scale). (b) The transmission through the cell obtained by momentarily turning off the current through the solenoids, again with the reference spectroscopy signal. The data were obtained with the unit having six layers of wire; the current increases from 1.00 to 2.75~A in steps of 0.25~A. }\label{fig:fig4}
\end{figure}

Figure~\ref{fig:fig4}  shows data obtained for the $^{87}$Rb $F=1 \rightarrow F^{'}$  transitions with 6 layers of wire.
At room temperature this is the transition with the smallest absorption, therefore we analyze the dependence of these spectra more closely.  Similar results for all the other transitions are analyzed in \cite{Danny}.

\begin{figure}[!ht]
\centering
\includegraphics[trim = 0mm 0mm 125mm 180mm, clip, scale=1]{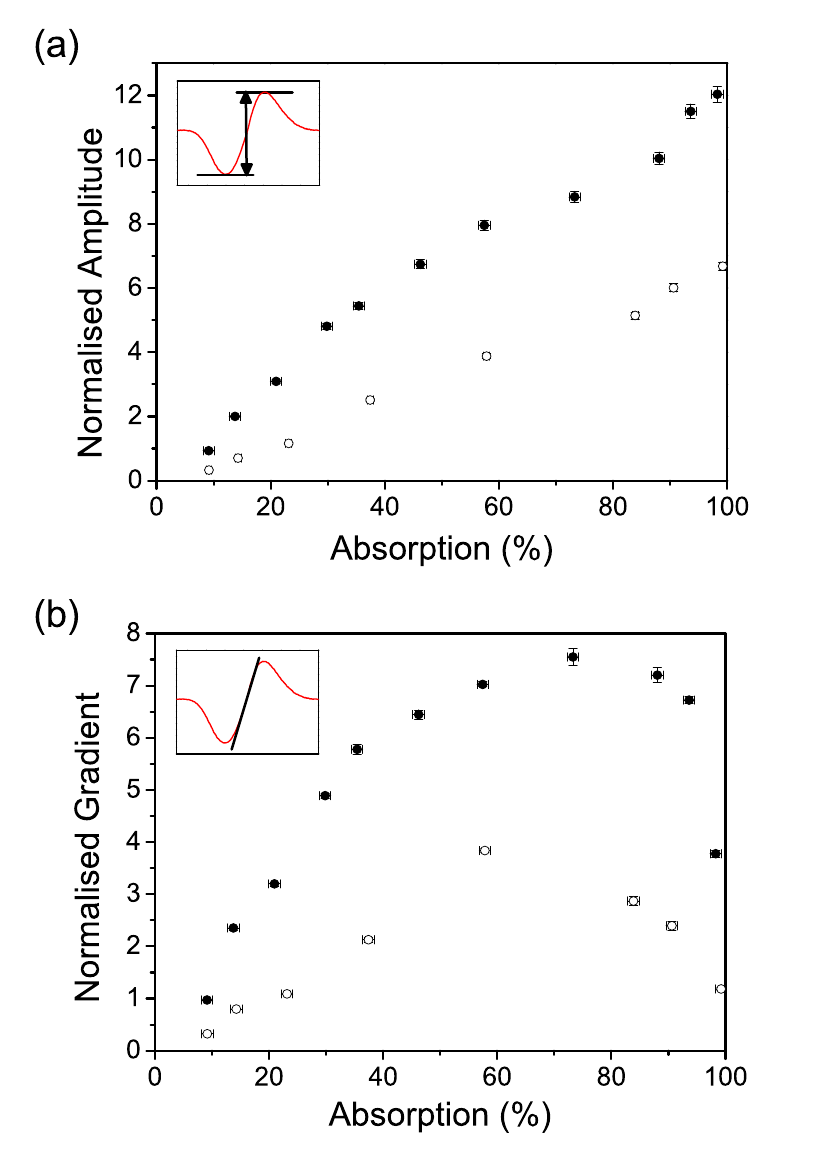}
\caption{DAVLL on the $^{87}$Rb $F=1 \rightarrow F^{'}$ transitions.
Parts (a) and (b) show the normalised amplitude and gradient, respectively, as a function of the line-center percentage absorption for 2 (open symbols) and 6 (solid symbols) layers of wire.}\label{fig:fig5}
\end{figure}

The data of figure~\ref{fig:fig5}   show the evolution of the  amplitude and gradient for the $^{87}$Rb $F=1 \rightarrow F^{'}$  transitions for different  \% absorption.  Data sets were taken with both six and two  layers of wire.  For increasing  current the amplitude and capture range grow monotonically.   However, once the
line-center absorption exceeds $\sim$75\%, the gradient falls dramatically. The reduction in gradient corresponds to the absorption saturating, and there being a range of laser frequencies over which there is little variation in the transmitted power.  As the frequency stability is dependent on the line-center slope of the error function, this implies that the optimum operating conditions correspond to a line-center absorption of $\sim$75\% - the small gain in amplitude for a more opaque cell is not enough to compensate for the significantly reduced  gradient.  The unit gave better lock signals (steeper gradient, larger amplitude) with six rather than two layers of wire as the magnetic field produced was closer to the optimal field \cite{Fred07} for the same enhancement in the absorption. For example, to obtain a line-center absorption of $\sim$75\% with two layers a current of 3.8~A is required yielding a 73~Gauss field, whereas with six layers a current of 2.25~A is required giving 124~Gauss. The latter is close to the $\sim$120~Gauss field required to optimize the signal gradient \cite{Fred07}, whereas the former is too small.

To facilitate a comparison of the  performance of the  compact unit, DAVLL signals were obtained with the
standard configuration with a 28~cm long water-cooled solenoid producing a field of 102~Gauss in a 5~cm long cell at room temperature. Table~\ref{max-summary}   summarizes the performance of this  set up. For  the weakest absorption line, the heated cell gave signal amplitudes which are larger by a factor of up to 12 and a gradient steeper by a factor up to 7.5.

\begin{table}
  \centering
\begin{tabular}{|l|c|c|c|c|}
\hline
 \rule[-1mm]{0mm}{5mm}& \multicolumn{2}{c|}{Standard} & \multicolumn{2}{c|}{Heated cell}\\ \cline{2-5}
 \rule{0mm}{4mm} Transition &  Amplitude & Gradient & Normalized  & Normalized  \\
 \rule[-2mm]{0mm}{4mm}  & (V) & (mV/MHz)& amplitude & gradient \\ \hline
 \rule[-1mm]{0mm}{5mm} $^{87}$Rb $F=2 \rightarrow
F^{'}$  & 0.99 $\pm$ 0.01 &  2.91 $\pm$ 0.05 & 6.16 $\pm$ 0.04 & 4.59 $\pm$ 0.06
\\ \hline
 \rule[-1mm]{0mm}{5mm} $^{87}$Rb $^{85}$Rb crossover  &  1.64 $\pm$ 0.01&   $-$4.42 $\pm$ 0.06  & 4.16 $\pm$ 0.07 & $-$3.87 $\pm$ 0.07 \\ \hline
 \rule[-1mm]{0mm}{5mm} $^{85}$Rb $F=3 \rightarrow F^{'}$  & 2.34 $\pm$ 0.01 &  7.74 $\pm$ 0.06& 3.85 $\pm$ 0.09 & 1.79 $\pm$ 0.08 \\ \hline
 \rule[-1mm]{0mm}{5mm} $^{85}$Rb $F=2 \rightarrow F^{'}$ & 1.71 $\pm$ 0.01& 4.18 $\pm$ 0.03  & 5.01 $\pm$ 0.05 & 4.49 $\pm$ 0.06 \\ \hline
 \rule[-1mm]{0mm}{5mm} $^{87}$Rb $F=1 \rightarrow F^{'}$ &  0.57 $\pm$ 0.01&1.62 $\pm$ 0.01  & 12.04 $\pm$ 0.02 & 7.55 $\pm$ 0.04 \\ \hline
\end{tabular}
  \caption{Summary of the  signal amplitude and gradient  for each transition with the standard DAVLL set up, and the factor by which the quantities are improved in the heated cell.}\label{max-summary}
\end{table}

For different atoms used in a DAVLL spectrometer, or even Rb with a different  cell length, it will be necessary to gather data empirically to ascertain  the optimum number of layers of solenoid wire.  However, we have shown above that the choice of number of layers is not a critical design constraint for compact heated cells, and a large improvement in signal amplitude and gradient relative to the room-temperature spectra can be generated.

An important feature of a DAVLL spectrometer is the stability of the lock point (the offset between the frequency of the signal zero crossing and the atomic resonance).  When used on a day-to-day basis in a laser-cooling experiment the
slow drift associated with permanent magnets has largely been eliminated with the unit presented here; the drift rate is so small that the quality of the lock is restricted by the laser stability.

In summary, we have presented the design and performance of a compact heated vapor cell unit for DAVLL which yields better stability and an order of magnitude improved performance for the $^{87}$Rb $F=1 \rightarrow F^{'}$ transition when compared to a standard configuration.

\section{Acknowledgements.}
This work is supported by EPSRC. SLC acknowledges the support
of the Royal Society. We thank Charles Adams and Matt Jones
for fruitful discussions. Victoria Greener provided the photograph of the assembled unit.

\bibliography{CompactDAVLL}

\end{document}